**Application of an interspecific competition model to predict the growth of *Aeromonas hydrophila* on fish surfaces during the refrigerated storage**

**Anwendung eines Konkurrenzmodells zwischen zwei Spezies zur Vorhersage des Wachstums von *Aeromonas hydrophila auf* Fischoberflächen während der Kühlphase**

Alessandro Giuffrida[1, *], Graziella Ziino[1], Davide Valenti[2], Giorgio Donato[1], Antonio Panebianco[1]

[1] *Dipartimento di Sanità Pubblica Veterinaria, Section of Inspection of Food of Animal Origin, Università di Messina, Viale Annunziata, 98168 Messina, Italy*

[2] *Dipartimento di Fisica e Tecnologie Relative and CNISM, Unità di Palermo, Group of Interdisciplinary Physics[#], Università di Palermo, Viale delle Scienze, I-90128 Palermo, Italy*

[*] Correspondence: Alessandro Giuffrida, Dipartimento di Sanità Pubblica Veterinaria, Polo Universitario dell'Annunziata, 98168 Messina – Italy. E-mail address: agiuffrida@unime.it Tel. +39 903503765; fax +39 903503937

**Abstract**

The growth of *Aeromonas hydrophila* and aerobic natural flora (APC) on gilthead seabream surfaces was evaluated during the refrigerated storage (21 days). The related growth curves were compared with those obtained by a conventional third order predictive model obtaining a low agreement between observed and predicted data (Root Mean Squared Error = 1.77 for *Aeromonas hydrophila* and 0.64 for APC). The Lotka-Volterra interspecific competition model was used in order to calculate the degree of interaction between the two bacterial populations ($\beta_{Ah/APC}$ and $\beta_{APC/Ah}$, respectively, the interspecific competition coefficients of APC on *Aeromonas hydrophila* and vice-versa). Afterwards, the Lotka-Volterra equations were applied as tertiary predictive model, taking into account, simultaneously, the environmental fluctuations and the bacterial interspecific competition. This approach allowed to obtain a best fitting to the observed mean growth curves with a Root Mean Squared Error of 0.09 for *Aeromonas hydrophila* and 0.28 for APC. Finally, authors carry out some considerations about the necessary use of competitive models in the context of the new trends in predictive microbiology.

**Zusammenfassung**

Das Wachstum von *Aeromonas hydrophila* und aerober natürlicher Flora (APC) auf der Oberfläche von Seebrassen wurde während der Kühlphase (21 Tage) ausgewertet. Die verwandten Wachstumskurven wurden mit solchen verglichen, die die durch ein konventionelles Vorhersagemodell dritter Ordnung ermittelt wurden. Letztere zeigen eine geringe Übereinstimmung zwischen vorhergesagten und beobachteten Daten (Standardabweichung = 1.77 für *Aeromonas hydrophila* und 0.64 für APC).

Das Lotka-Volterra Konkurrenzmodell zwischen zwei Spezies wurde zur Berechnung des Grades der Interaktion zwischen den beiden Bakterienpopulationen benutzt (mit den Konkurrenzkoeffizienten für APC *gegenüber Aeromonas hydrophila* und umgekehrt *βAh/APC* bzw. *βAPC/Ah* ). Danach wurden die Lotka-Volterra Gleichungen als tertiäres Vorhersagemodell angewandt, wobei gleichzeitig Umwelteinflüsse und die Konkurrenz zwischen den Bakterienspezies berücksichtigt wurden.

Dieser Ansatz erlaubte einen Fit and die beobachteten mittleren Wachstumskurven mit einer Standardabweichung von 0.25 für *Aeromonas hydrophila und 0.28 für APC.*

Zuletzt folgen einige Betrachtungen zur Anwendung von Konkurrenzmodellen im Kontext neuer Entwicklungen zur Vorhersage in der Mikrobiologie.

**Schlüsselwörter:** *Aeromonas hydrophila*, Goldbrasse, Vorhersagemodell, bakterielle Konkurrenz

1. Introduction

*Aeromonas hydrophila* (*Ah*) is an emerging foodborne disease agent, widely distributed in the environment. Recently, the genus *Aeromonas* has been classified within the family *Aeromonadaceae* and consists of 14 different confirmed species, one of which is *Ah* (Joseph and Carnahan, 2000). It is well known that the microorganism is cause of several disease conditions in cold (fish, reptiles, amphibians) and warm-blooded (mammals and birds) animals as well as of a zoonotic disease (Daskalov, 2006). *Ah* is widely spread in waters, water habitants, and many food products (seafood, shellfish, raw foods of animal origin like poultry, ground meat, raw milk, and raw vegetables) (Buchanan and Palumbo, 1985; Daskalov, 2006; Fricker and Tompsett, 1989; Gobat and Jemmi, 1993; Krovacek et al., 1992; Nishikawa and Kishi, 1988). With regard to foods of animal origin, according to Kumar et al. (2000) and Neyts et al. (2000)*,* seafood products are more frequently contaminated by *Ah* in consequence of the wide diffusion in the aquatic environment and the ability to grow at cold temperatures. Furthermore, since *Ah* is an important agent of several freshwater (Aoki, 1999) and marine (Balebona et al., 1998; Zorrilla et al., 2003) reared fish diseases, its spread in aquaculture environment could be a significant public health concern (Daskalov, 2006; Giuffrida, 2003). However, the above food safety implications are strictly related to the pathogenicity and virulence of the strain as well as to the bacterial concentration which *Ah* is able to reach during the storage. In the case of fish, the main *Ah* growth during the storage, like for several other microorganisms, occurs on the skin and the gills which are considered the most important source of spoilage and pathogen bacteria (Giuffrida, 2003, Kumar et al., 2000).

The growth ability of *Ah* was studied by several authors (Palumbo et al., 1985; Palumbo et al., 1991; Palumbo et al., 1992; Palumbo et al., 1996; Pin et al., 2004) which developed some predictive modelling techniques to better understand the potential behaviour of the bacteria under different kinds of food storage conditions. Nevertheless, as for other bacteria, some authors stressed the discrepancy between the growth in broth and in food supposing the natural competitive flora was the cause of this discrepancy (Gill et al., 1997, Palumbo et al., 1988, Pin et al., 2004). The competition of natural flora is a complex issue in the modelling of microbial evolution and it was studied by several authors. Pin and Baranyi (1998) studied the interactions of some groups of spoilage organisms that can be usually found in refrigerated meat, quantifying the inhibition exerted by a specific spoilage group (*Pseudomonas* spp. / *Shewanella* spp.) on each one of the other groups, in the range of temperature 2–11°C and pH 5.2–6.4. Giménez and Dalgaard (2004) modelled the simultaneous growth of *Listeria monocytogenes* and spoilage microflora in cold-smoked salmon by introducing into a simple differential equation an additional term which gives account for the interaction between *Listeria monocytogenes* and spoilage microflora, so that they inhibit each other to the same extent that they inhibit their own growth.

Vereecken et al. (2003) introduced a new methodology for modelling the bacterial competition as a function of lactic acid production. This element was incorporated into the well known logistic model of Baranyi and Roberts (1994) substituting the inhibition function $\mu_N(N)=(1-N/N_{max})$ with the function $\mu_P(P)$. This term depends on the metabolic product $P(m)$, that symbolizes the total lactic acid concentration (mechanistic inhibition function). A further development of the study of Vereecken et al. (2003) has been recently carried out by Van Impe et al. (2005), with the introduction of the $\mu_S(S)$ factor which describes the influence of the phenomenon of exhaustion of a substrate $(S)$ on the microbial evolution.

Another interesting approaches to the bacterial competition modelling is based on the Lotka-Volterra competition model which provides a basic equation for the population growth of two interacting species. A prototype model structure for mixed microbial populations in food products was proposed by Dens et al. (1999) which combined the advantages of the Lotka-Volterra model for two-species competition with those of the model by Baranyi & Roberts (1994) as a classical predictive growth model. Powell et al. (2004) used the aforesaid model in order to interpret some empirical results for *Escherichia coli* O157:H7 in ground beef and they showed that the seemingly incongruous data were consistent with the interspecific competition model. In this regard, it is important to stress that the Lotka-Volterra model can not be considered exactly like a simple "primary" predictive model (Whiting and Buchanan, 1993) since it can not be used in order to describe, directly and simultaneously, the observed microbial evolution of two competitive specie as a function of time. For example, in one of the simulated scenarios of the study of Powell at al. (2004), the authors used the growth rate for *Escherichia coli* O157:H7, already found by Walls and Scott (1996). Afterwards, holding other factors constant, each competition model term was varied until the Theoretically Maximum Population Densities (TMD) reached the observed Maximum Population Densities (MPD). This implies, therefore, that, fixing the growth parameters of each bacterial species (such as maximum specific growth rate and physiological state of the species), the Lotka-Volterra model could be used in order to calculate the competition terms ($\beta_{12}$ and $\beta_{21}$ of the following equations 1a and 1c) by fitting the predictive behaviours of two competitive species to the observed growth curves. However, in this case it would be necessary to use growth parameters obtained from mono-cultures experimental data in order to consider only once the effect of interspecific bacterial competition. The Lotka-Volterra competition model, as proposed by Dens et al. (1999) and Powell et al. (2004) is represented by the following set of differential equations.

$$\frac{dN_1}{dt} = \mu_1^{max} N_1 \frac{Q_1}{1+Q_1}\left(1 - \frac{N_1 + \beta_{12} N_2}{N_1^{max}}\right) \tag{1a}$$

$$\frac{dQ_1}{dt} = \mu_1^{max} Q_1 \tag{1b}$$

$$\frac{dN_2}{dt} = \mu_2^{max} N_2 \frac{Q_2}{1+Q_2}\left(1 - \frac{N_2 + \beta_{21} N_1}{N_2^{max}}\right) \tag{1c}$$

$$\frac{dQ_2}{dt} = \mu_2^{max} Q_2 \tag{1d}$$

where $N_1$ and $N_2$ are, respectively, the population densities of competitive species at time t, $\mu_1^{max}$ and $\mu_2^{max}$ are the maximum specific growth rates of both species, $N_1^{max}$ and $N_2^{max}$ are the theoretically maximum population densities under monospecific growth conditions, $\beta_{12}$ and $\beta_{21}$ are, respectively, the interspecific competition terms of *species₂* on *species₁* and vice-versa; $Q_1$ and $Q_2$ represent, respectively, the physiological state of two bacterial populations.

By fixing the values of the interspecific competition terms for two bacterial populations (*species₁* and *species₂*) it is also possible to incorporate into the Lotka-Volterra equations a "secondary" model which relates the maximum specific growth rates of both species to the environmental parameters (T, pH, aw, $CO_2$ and $O_2$ percentage, etc.). In this way the model will work as a "tertiary" predictive model (Whiting and Buchanan, 1993) which is able to take into account, simultaneously, not only the environmental dynamics, like a conventional tertiary predictive model, but also the microbial community competitions.

In this work we analysed the growth of *Ah* on reared gilthead seabream (*Sparus aurata*) during the refrigerated storage, comparing the observed behaviour with the predicted growth by a conventional tertiary model. Afterwards, we presented a practical application of the aforementioned theoretical approach for the evaluation of: i) the interaction intensity between *Ah* and the aerobic natural flora (Aerobic Plate Count, *APC*); ii) the suitability of the incorporation of secondary models (polynomial equations) into the Lotka-Volterra equations in order to obtain a better fit of the observed *Ah* behaviour on gilthead seabream surfaces during the storage, taking also into account the fluctuations of storage temperature.

**2. Materials and Methods**

*2.1. Samples and bacteriological analysis*

Number 95 specimens of *Sparus aurata*, which resulted positive for *Ah* in the context of a wider research concerning the quality of marine reared fish, were analysed in order to a quantitative determination of *Ah* and *APC* on skin and gills. In particular, the evaluations were carried out with regard to n. 63 fish with naturally contaminated skin by *Ah* and n. 32 fish with naturally contaminated gills by *Ah*. Samples were collected at time 0 and after 72, 168, 240, 336, 408, 504 hours of refrigerated storage. The storage temperature was monitored using FT-800/SYSTEM temperature data-loggers (Econorma s.a.s., Vendemiano, Italy) placed on fish surface. FT-800/SYSTEM loggers were set to record the temperature every 1.2 min.

For the *Ah* enumeration, the skin and gills were extracted with sterile instruments; tenfold dilutions in 0.1% Phosphate Buffered Broth (Oxoid, Basingstoke, UK) were performed according to ISO 6887-1:1999 and 0.1 ml of each dilution was plated in duplicated onto Aeromonas Agar (AA) (Oxoid) and GSP Agar (Merk KGaA, Darmstadt, Germany), incubated at 30°C for 24-48 hours. For a better discrimination between presumptive *Aeromonas* spp. and *Pseudomonas* spp., plates of AA and GSP Agar were incubated both aerobically and anaerobically (anaerobic jars with a gas-generating kit; BioMérieux, Marcy l'Etoile, France). Green colonies in AA, and yellow colonies in GSP agar, were counted as presumptive *Aeromonas spp.*. The 50% of presumptive *Aeromonas* spp. colonies were identified by morphological and biochemical (API 20E system, BioMérieux) tests as well as, in the context of the above wider research project, by PCR techniques whose methodologies and results are not treated in the present paper. For APC, 1 ml of each tenfold dilution was plated in three Petri dishes, covered by Plate Count Agar (Oxoid) and incubated at 25°C for 3 days.

*2.2. Analysis of growth curves*

N. 10 *Ah* growth curves (n. 5 growth curves for the skin and n. 5 for the gills) and as many for *APC* were obtained by expressing the number of colony forming units (CFU) as decimal logarithm (Log CFU/g); for the aerobic natural flora enumeration, the *Ah* CFU number was previously subtracted from the APC. The growth curves were analysed by the well known following differential equations of Baranyi and Roberts (1994)

$$\frac{dN}{dt} = \mu_{max} N \frac{Q}{1+Q}\left(1-\frac{N}{N_{max}}\right) \qquad (2a)$$

$$\frac{dQ}{dt} = \mu_{max} Q \qquad (2b)$$

in order to calculate the main growth parameters, where $N$ is the population densities at time t, $\mu_{max}$ is the maximum specific growth rate, $N_{max}$ is the theoretically maximum population densities and $Q$ is the physiological state of the species which allows to obtain the Lag-time, since

$$Lag\_time = \frac{-\ln \alpha}{\mu_{max}} \qquad (3)$$

and

$$\alpha = \frac{Q}{1+Q} \qquad (4)$$

The differential equations 2a and 2b were solved, for each growth curve, by Runge-Kutta method and $\mu_{max}$, $Q$ and $N_{max}$ were obtained by using the Solver of Microsoft Excel (Office 2003 package, Microsoft Windows corporation®). Observed and fitted growth curves were statistically analysed by using the Root Mean Squared Error (RMSE).

*2.3. Conventional prediction of A. hydrophila and APC growth*

The prediction of *Ah* and APC growth was carried out by using the conventional structure of a tertiary predictive model which combines primary and secondary models (Whiting and Buchanan, 1993). In particular, for *Ah*, the polynomial equation 5 (secondary model) proposed by Pin et al. (2004) and obtained from mono-cultures experimental data, was substituted into the model of Baranyi and Roberts (1994) (Eqs. 2a-b).

$$\text{Ln}(\mu_{Ah}^{max}) = b_0 + (T)*b_1 + (T)^2*b_2 + (T*CO_2)*b_3 + (T*pH)*b_4 + (pH*O_2)*b_5 \qquad (5)$$

where $b_0$ = -4.5; $b_1$=-829; $b_2$=-0.0151; $b_3$=-0.00122; $b_4$=0.184; $b_5$=-0.00114.

In the case of *APC* prediction, the secondary model was constructed by acquiring several growth curves from ComBase (Institute of Food Research – UK; http://wyndmoor.arserrc.gov/combase/); the logarithms of observed $\mu_{max}$ were regressed to the respective temperature and pH values and a stepwise procedure was used to remove those coefficients that did not contribute significantly to the model (StatTools add-in for Microsoft Excel; Palisade Corporation®, 2003). In this way the coefficients $b_0 - b_2$ of the following equation 6 were obtained.

$$\text{Ln}(\mu_{APC}^{\max}) = b_0 + (pH)*b_1 + (T)^2*b_2 \qquad (6)$$

where $b_0$ = -2.5050; $b_1$=-0.2267; $b_2$=0.0072.

The equation 6 was substituted into the model of Baranyi and Roberts (1994) in order to construct the third order model for *APC*.

Both equation systems were solved numerically, by the fourth-order Runge–Kutta method, to obtain predictions for the bacterial concentration during time-dependent temperature profiles, while the other environment parameters were maintained constant (pH=7.0; $CO_2$%=1.0; $O_2$%=20). The initial bacterial concentrations were taken as the observed inoculum and, for the initial value of $Q_0$, the procedure of Baranyi et al. (1995) was followed for both populations. The $N_{max}$ values for *Ah* and *APC* (respectively Log 7.26 CFU/g and Log 9.5 CFU/g) were obtained from the available growth curves of the ComBase online database.

The *Ah* and *APC* predicted growth curves were compared to the mean growth curves of the observed values of both populations and the differences were statistically analysed through the Root Mean Squared Error (RMSE).

*2.4. Lotka-Volterra competition model*

The following Lotka-Volterra competition model (Eqs. 7a-d) was used in order to analyse the interaction between *Ah* and *APC* and to obtain a better fitting of the theoretical results to the observed *Ah* growth on gilthead sea bream surfaces.

$$\frac{dN_{Ah}}{dt} = \mu_{Ah}^{\max} \, N_{Ah} \frac{Q_{Ah}}{1+Q_{Ah}} \left(1 - \frac{N_{Ah} + \beta_{Ah/APC} N_{APC}}{N_{Ah}^{\max}}\right) \qquad (7a)$$

$$\frac{dQ_{Ah}}{dt} = \mu_{Ah}^{\max} \, Q_{Ah} \qquad (7b)$$

$$\frac{dN_{APC}}{dt} = \mu_{APC}^{\max} \, N_{APC} \frac{Q_{APC}}{1+Q_{APC}} \left(1 - \frac{N_{APC} + \beta_{APC/Ah} N_{Ah}}{N_{APC}^{\max}}\right) \qquad (7c)$$

$$\frac{dQ_{APC}}{dt} = \mu_{APC}^{\max} \, Q_{APC} \qquad (7d)$$

where $N_{Ah}$ and $N_{APC}$ are, respectively, the population densities of *Ah* and *APC* at time t, $\mu_{Ah}^{\max}$ and $\mu_{APC}^{\max}$ are the maximum specific growth rates of both populations, $N_{Ah}^{\max}$ and $N_{APC}^{\max}$ are the theoretically maximum population densities under monospecific growth conditions, $β_{Ah/APC}$ and $β_{APC/Ah}$ are, respectively, the interspecific competition terms of APC on Ah and vice-versa; $Q_{Ah}$ and $Q_{APC}$ represent, respectively, the physiological state of two bacterial populations.

Analogously to the above conventional third order models, equations 5 and 6 were substituted, respectively, into equations 7a-b and 7c-d allowing the instantaneously modification of growth rate during time-dependent temperature profiles; also in this case the other environment parameters were maintained constant (pH=7.0; $CO_2$%=1.0; $O_2$%=20). $N_0$, $Q_0$ and $N_{max}$ values of both population were the same of those used for the construction of the conventional third order model.

The system was solved numerically by the fourth-order Runge–Kutta method and, at the same time, the Solver of Microsoft Excel (Office 2003 package, Microsoft Windows corporation®) was used in order to fit the predicted Ah and APC behaviours to the mean growth curves of the observed values, by modifying both β-terms ($β_{Ah/APC}$ and $β_{APC/Ah}$, respectively in equations 7a and 7c).

Also in this case, the Ah and APC predicted growth curves were compared to the mean growth curves of the observed values of both populations and the differences were statistically analysed through the Root Mean Squared Error (RMSE).

Note that this approach allows to consider only once the effect of interspecific bacterial competition (β-term) since, as explained in the above section, the other parameters derives from monoculture experimental data.

## 3. Results and discussion

The observed mean growth curves for Ah and APC as well as the population behaviours predicted by the conventional tertiary models are showed in Figure 1; in the same figure is also reported the mean recorded temperature during the refrigerated storage. The mean growth parameters estimated for the Ah growth curves on fish surfaces show a very slow increase with a Lag-time of 241.30±46.60 hours and a $μ_{max}$ of 0.0276±0.0078. The estimated mean growth parameters of APC show a similar behaviour of this bacterial population with a Lag-time of 190.12±81.24 hours and a $μ_{max}$ of 0.0179±0.009 . As Figure 1 graphically shows, the predictions with the conventional tertiary model do not agree with the observed data since the RMSE values were 1.77 for Ah and 0.64 for APC.

Concerning the application of the Lotka-Volterra interspecific competition model, the results are summarised in Table 1 which shows, in first instance, as the terms $β_{Ah/APC}$ (interaction of APC on Ah) was higher than $β_{APC/Ah}$ (interaction of Ah on APC), at least for the experimental conditions of this work. This aspect is easily explainable since APC should be presumably constituted by *Pseudomonas* spp., *Shewanella* spp., *Acinetobacter* spp. and other fish spoilage bacteria which can have an interactive effect by competing for the nutrients as well as by a siderophore-mediated competition for iron. This mechanism, for example, was demonstrated for *Pseudomonas* spp. against *Staphylococcus* spp, *Escherichia coli* and *Aeromonas hydrophila* (Vachée et al., 1997). However, the *β* term values (<1.0) indicate a regime of coexistence between species (Dens et al., 1999; Spagnolo et al., 2004; Valenti et al., 2004) inducing to suppose that this competitive activity is not comparable, for example, to the Lactic Acid Bacteria competition for several foodborne disease agents.

Figure 2 shows the predictions obtained by the incorporation of equations 5 and 6 into the Lotka-Volterra model as well as the mean of the observed data and the mean temperature records. In this case the model allowed to take into account, simultaneously, the environmental influences (temperature fluctuations) on bacterial growth and the interspecific competitions (*β* terms; Tab. 1) which are concentration dependent. As Figure 2 and Table 2 (RMSE values) show, this approach produced a better fitting of predicted to observed data, especially concerning the reaching of Maximum Population Density. It is important to stress that the proposed model allows to express the behaviour of two competitive bacterial populations as the results of two opposed forces: the former is represented by the growth rate of each population which derives from monoculture experimental data; the latter is the interspecific competition term (*β*) of each population on the other one which is related to the bacterial growth.

Finally, we can conclude that the proposed interspecific competition model represent a good solution in order to consider, at the same time, the complexity of the food substrate during the storage (fluctuating environmental condition) and the interspecific bacterial interactions. On the contrary, conventional predictive tertiary models which reproduce the mono-specific bacterial growth in food where the natural flora has a high increase during the storage, could produce incongruous data. This matter would be considered by the new trends in predictive microbiology since, as we showed, the bacterial interspecific competition can significantly affect the population behaviour.

It is evident that the limit of the present work is represented by the considering *APC* as a single bacterial population since it is well known that it includes several kinds of aerobic bacterial species. However, the aim of this work was to characterise the *Aeromonas hydrophila* behaviours on gilthead seabream surfaces, taking into account the competitive effect of all other bacterial species. On the contrary, the eventual characterisation of the competitive effect of only one spoilage aerobic species (e.g. *Pseudomonas* spp.) on *Aeromonas hydrophila* growth could produce a result with a minor applicative meaning. In this regard, a further development of the present study could consider the modelling of several competitive species (more than two) as theoretically proposed by Fiasconaro et al. (2006).

*Acknowledgements*. *The financial support of Minister of Instruction and Scientific Research is gratefully acknowledged (Project code: PRIN 2004079923_001; 2004)*

**TABLE 1:** Estimated parameters by using the Lotka-Volterra interspecific competition model. The Root Mean Squared Error (RMSE) values concern the comparison between the observed data and the Lotka-Volterra model estimations

|  | *Aeromonas hydrophila* | **Aerobic Plate Count** |
|---|---|---|
| *$\beta_{Ah/APC}$** | | 0.3372 |
| *$\beta_{APC/Ah}$**** | | 0.2420 |
| **Lag time** | 218.40 | 199.46 |
| **Nmax** | 4.4096 | 8.3953 |
| **RMSE** | 0.0912 | 0.2833 |

*\* Interspecific competition term of APC on Ah*

*\*\* Interspecific competition term of Ah on APC*

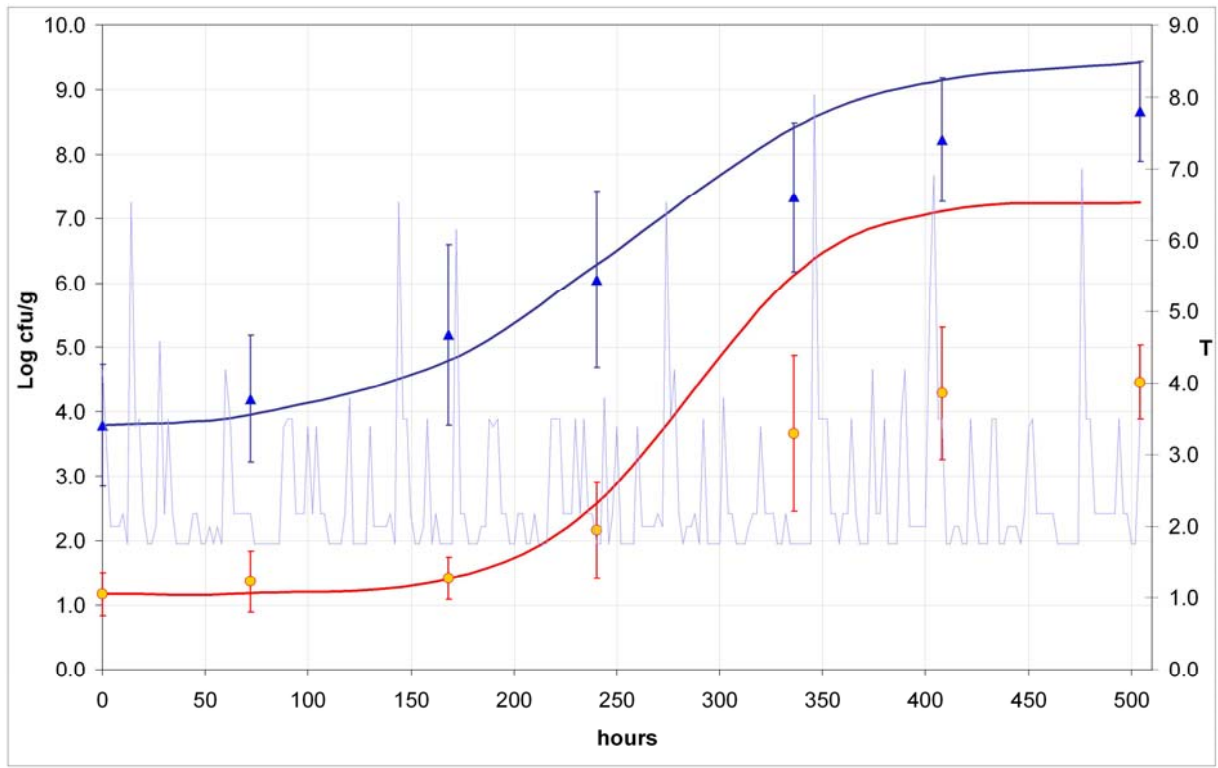

**Fig. 1.**

Observed behaviours of *Aeromonas hydrophila* (○) and Aerobic Plate Count (▲) during the refrigerated storage at fluctuating temperature (—) of gilthead seabream. Red and blue straight lines indicate, respectively, the predicted growth of *Aeromonas hydrophila* (—) and Aerobic Plate Count (—) using the conventional third order model approach.

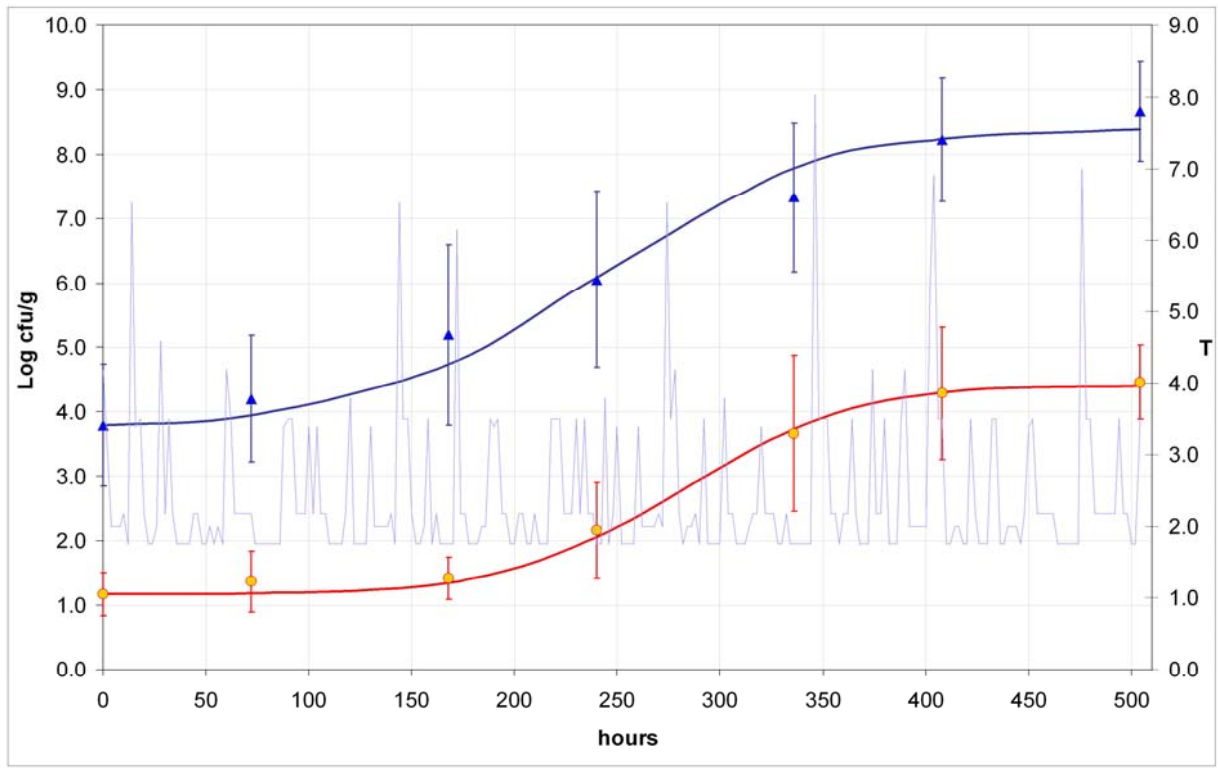

**Fig. 2.**

Observed behaviours of *Aeromonas hydrophila* (○) and Aerobic Plate Count (▲) during the refrigerated storage at fluctuating temperature (—) of gilthead seabream. Red and blue straight lines indicate, respectively, the predicted growth of *Aeromonas hydrophila* (—) and Aerobic Plate Count (—) using the Lotka-Volterra interspecific competition model.